\documentclass{article}

\usepackage{microtype}
\usepackage{graphicx}
\usepackage{booktabs} 

\usepackage[accepted]{icml2024}

\usepackage{amsmath}
\usepackage{amssymb}
\usepackage{mathtools}
\usepackage{amsthm}
\usepackage{subcaption}

\usepackage{multirow}
\usepackage{comment} 
\usepackage{enumitem}

\usepackage{times}
\usepackage{epsfig}
\usepackage{graphicx}

\usepackage[bb=dsserif]{mathalpha}
\usepackage{bm}

\usepackage{booktabs,colortbl,tabularx}
\usepackage{pifont}%

\usepackage{mathtools}
\usepackage{commath}
\usepackage{algorithm}

\usepackage{multirow}
\usepackage{comment} 
\usepackage{enumitem}

\definecolor{Gray}{gray}{0.9}

\newcommand{\cmark}{\ding{51}}%
\newcommand{\xmark}{\ding{55}}%

\definecolor{battleshipgrey}{rgb}{0.52, 0.52, 0.51}

\usepackage{xspace}
\newcommand{\method}{\textsc{T2AV}\xspace}
\newcommand{\benchmark}{\textsc{T2AV-Bench}\xspace}

\usepackage[capitalize,noabbrev]{cleveref}

\theoremstyle{plain}

\theoremstyle{definition}

\theoremstyle{remark}

\icmltitlerunning{Text-to-Audio Generation Synchronized with Videos}

\begin{document}

\twocolumn[
\icmltitle{Text-to-Audio Generation Synchronized with Videos}



\icmlsetsymbol{equal}{*}

\begin{icmlauthorlist}
\icmlauthor{Shentong Mo}{cmu}
\icmlauthor{Jing Shi}{ar}
\icmlauthor{Yapeng Tian}{utd}
\end{icmlauthorlist}

\icmlaffiliation{cmu}{Carnegie Mellon University}
\icmlaffiliation{ar}{Adobe Research}
\icmlaffiliation{utd}{University of Texas at Dallas}

\icmlcorrespondingauthor{Shentong Mo}{shentongmo@gmail.com}

\icmlkeywords{Machine Learning, ICML}

\vskip 0.3in
]



\printAffiliationsAndNotice{}  

\begin{abstract}

In recent times, the focus on text-to-audio (TTA) generation has intensified, as researchers strive to synthesize audio from textual descriptions. 
However, most existing methods, though leveraging latent diffusion models to learn the correlation between audio and text embeddings, fall short when it comes to maintaining a seamless synchronization between the produced audio and its video. 
This often results in discernible audio-visual mismatches. 
To bridge this gap, we introduce a groundbreaking benchmark for Text-to-Audio generation that aligns with Videos, named \benchmark. 
This benchmark distinguishes itself with three novel metrics dedicated to evaluating visual alignment and temporal consistency. 
To complement this, we also present a simple yet effective video-aligned TTA generation model, namely \method. 
Moving beyond traditional methods, \method refines the latent diffusion approach by integrating visual-aligned text embeddings as its conditional foundation. 
It employs a temporal multi-head attention transformer to extract and understand temporal nuances from video data, a feat amplified by our Audio-Visual ControlNet that adeptly merges temporal visual representations with text embeddings. 
Further enhancing this integration, we weave in a contrastive learning objective, designed to ensure that the visual-aligned text embeddings resonate closely with the audio features. 
Extensive evaluations on the AudioCaps and \benchmark demonstrate that our \method sets a new standard for video-aligned TTA generation in ensuring visual alignment and temporal consistency.

\end{abstract}

\section{Introduction}

The realm of audio signal processing is continually expanding, and one of its most promising offshoots is text-to-audio (TTA) generation. 
This avenue seeks to address a compelling query: \textit{is it feasible to craft high-dimensional audio signals that are not only sonically rich but also contextually attuned to their textual precursors? }
Recent times have witnessed a surge of investigative endeavors into this realm, with researchers exploring the power of denoising diffusion probabilistic models (DDPMs) as evidenced by innovations like DiffSound~\citep{yang2022diffsound} and AudioGen~\citep{kreuk2023audiogen}. 
Building on this momentum, AudioLDM~\citep{liu2023audioldm} set a new trajectory by synergizing latent diffusion models with the contrastive prowess of language-audio pre-training (CLAP)~\citep{laionclap2023}, rooting their strategy in the bedrock of text embeddings.
The resultant audio landscapes were not just audibly impressive but also contextually tethered to their textual origins.

\begin{figure}
\centering
\includegraphics[width=1.0\linewidth]{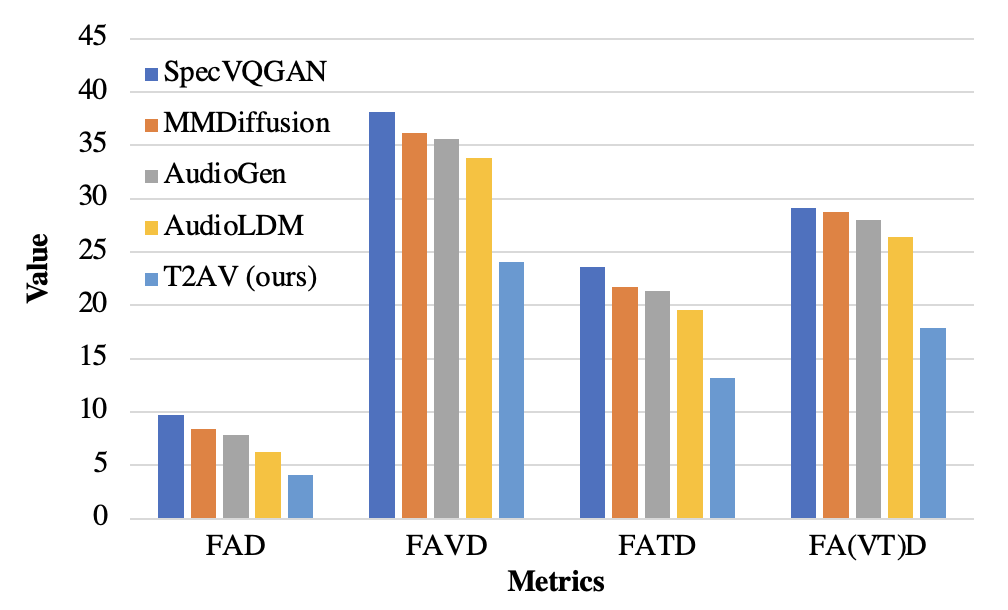}
\vspace{-7mm}
\caption{{\bf Comparison of our \method with state-of-the-art methods} on the proposed \benchmark in terms of FAD, FAVD, FATD, and FA(VT)D for video-aligned text-to-audio generation. 
Our method significantly outperforms previous baselines in terms of all metrics (lower is better).}
\vspace{-5mm}
\label{fig: title_img}
\end{figure}

Although these methods achieve impressive performance in generating plausible sounds, they ignored the synchronization between generated audio and visual content in videos, resulting in misaligned audio and video frames. 
For instance, the model might generate a train horn sound for ten seconds, even when no train is visible in the frame.
The main challenge is that sounds are naturally aligned with frames in natural videos. 
This inspires us to learn video-aligned semantics for each text prompt from the video to guide text-to-audio generation. 
To address the problem, our key idea is to capture video-aligned text representations for updating text embeddings as the condition, which differs from existing DDPMs and LDMs on TTA generation.

To address this challenge, we present the \benchmark benchmark for Text-to-Audio generation aligned with Videos to ensure synchronization with video content. 
Meanwhile, inspired by Frechet Inception Distance in the image domain, we present three novel metrics including Frechet Audio-Visual Distance, Frechet Audio-Visual Distance, and Frechet Audio-(Video-Text) Distance for the evaluation of the quality of our generated audio regarding visual alignment and temporal consistency.

Beyond this benchmark, we also introduce \method, a simple yet effective method based on a latent diffusion model that can learn visual-aligned text semantics as guidance for video-aligned TTA generation. 
Specifically, our \method leverages visual-aligned contrastive language-audio pre-training to capture the alignment between the textual and visual features at spatial and temporal levels corresponding to the paired audio.
Then, we introduce Audio-Visual ControlNet based on a temporal multi-head attention transformer to extract text embeddings with visual-aligned semantics as the condition for latent diffusion models.
Compared to previous TTA baselines, our new framework can support a flexible number of condition operators in Audio-Visual ControlNet and shows the effectiveness of learning video-aligned semantics for generating high-fidelity audio given input captions

Empirical experiments on AudioCaps benchmarks and our \benchmark comprehensively demonstrate the state-of-the-art performance against previous text-to-audio generation baselines. 
In addition, qualitative visualizations of target-generated video results showcase the effectiveness of our \method in generating high-fidelity audio aligned with videos. 
Extensive ablation studies also validate the importance of visual-aligned language-audio pre-training and Audio-Visual ControlNet in learning temporal-aware representations for maintaining visual alignment and temporal consistency.
We also demonstrate the importance of the training data scale and latent diffusion tuning in video-aligned text-to-audio generation.

In summation, our contributions can be summarized as:
\begin{itemize}
\item We present a novel benchmark for Text-to-Audio generation aligned with Video, namely \benchmark, complemented by three new metrics focusing on visual coherence and temporal synchronicity.
\item We introduce a simple yet effective approach called \method, a new latent diffusion model that symbiotically integrates temporal video representations conditioned by the proposed Audio-Visual ControlNet.
\item Extensive experiments comprehensively demonstrate 
the state-of-the-art superiority of our \method over previous baselines on text-to-audio generation outputs with visual alignment and temporal consistency.
\end{itemize}

\section{Related Work}

\noindent{\textbf{Diffusion Models.}}
Diffusion models have been demonstrated to be effective in many generative tasks, such as image generation~\citep{saharia2022photorealistic}, image restoration~\citep{saharia2021image}, speech generation~\citep{kong2021diffwave}, and video generation~\citep{ho2022imagen}.
Typically, denoising diffusion probabilistic models (DDPMs)~\citep{ho2020denoising,song2021scorebased} utilized a forward noising process that gradually adds Gaussian noise to images and trained a reverse process that inverts the forward process.
Unlike them, we apply latent diffusion models (LDMs) on audio embeddings to generate visually aligned and temporally consistent sounds based on text descriptions.

\noindent{\textbf{Audio-Visual Learning.}}
Audio-visual learning has been explored in many previous works~\citep{aytar2016soundnet,owens2016ambient,Arandjelovic2017look,korbar2018cooperative,Senocak2018learning,zhao2018the,zhao2019the,Gan2020music,Morgado2020learning,Morgado2021robust,Morgado2021audio,mo2022semantic,mo2022benchmarking,mo2023diffava,mo2023oneavm,mo2023deepavfusion,mo2023class,pian2023audiovisual} to capture the audio-visual alignment between two distinct modalities in videos.
Such cross-modal correspondences are beneficial for many audio-visual tasks, such as audio-event localization~\citep{tian2018ave,wu2019dual,lin2019dual}, audio-visual parsing~\citep{tian2020unified,wu2021explore,lin2021exploring,mo2022multimodal}, audio-visual spatialization \& localization~\citep{Morgado2018selfsupervised,Morgado2020learning,mo2022EZVSL,mo2022SLAVC,mo2023audiovisual,mo2023avsam,mo2023weaklysupervised}, and visual-to-sound generation~\cite{zhou2018visual,iashin2021taming,du2023conditional}. While the task of visual-to-sound generation is relevant to our problem, our primary focus is on learning discriminative cross-modal representations for visual-guided text-to-sound generation, a more challenging endeavor than the aforementioned tasks.

\newcommand{\favd}{
\begin{tabular}{ccc}
    \toprule
    \bf True Pairs & \bf False Pairs & \bf FAVD ($\downarrow$) \\
    \midrule
    \rowcolor{blue!10}
    500 & 0 & \bf 12.85 \\
    0 & 500 & 58.62 \\
    500 & 500 & 36.72 \\
    500 & 1000 & 49.51 \\
    1000 & 500 & 29.18 \\
    \bottomrule
\end{tabular}
}

\newcommand{\fatd}{
\begin{tabular}{ccc}
    \toprule
    \bf True Pairs & \bf False Pairs & \bf FATD ($\downarrow$) \\
    \midrule
    \rowcolor{blue!10}
    500 & 0 & \bf 5.27 \\
    0 & 500 & 38.39 \\
    500 & 500 & 22.58 \\
    500 & 1000 & 30.39 \\
    1000 & 500 & 16.72 \\
    \bottomrule
\end{tabular}
}

\newcommand{\favtd}{
\begin{tabular}{ccc}
    \toprule
    \bf True Pairs & \bf False Pairs & \bf FA(VT)D ($\downarrow$) \\
    \midrule
    \rowcolor{blue!10}
    500 & 0 & \bf 8.75 \\
    0 & 500 & 42.19 \\
    500 & 500 & 26.47 \\
    500 & 1000 & 33.98 \\
    1000 & 500 & 21.33 \\
    \bottomrule
\end{tabular}
}

\begin{table*}[!htb]
    \centering
    \begin{subtable}{0.3\textwidth}
        \centering
        \resizebox{\linewidth}{!}{\favd}
        \caption{FAVD.}
       \label{tab: exp_favd}
    \end{subtable}
    \quad
    \begin{subtable}{0.3\textwidth}
        \centering
        \resizebox{\linewidth}{!}{\fatd}
        \caption{FATD}
        \label{tab: exp_fatd}
    \end{subtable}
    \quad
    \begin{subtable}{0.33\textwidth}
        \centering
        \resizebox{\linewidth}{!}{\favtd}
        \caption{FA(VT)D}
        \label{tab: exp_favtd}
    \end{subtable}
    \vspace{-1em}
    \caption{{\bf Visual alignment validation} of all metrics in our \benchmark.}
        \label{tab: exp_validation}
     \vspace{-0.5em}
\end{table*}

\newcommand{\favdc}{
\begin{tabular}{ccc}
    \toprule
    \bf True Pairs & \bf False Pairs & \bf FAVD ($\downarrow$) \\
    \midrule
    \rowcolor{blue!10}
    500 & 0 & \bf 12.85 \\
    0 & 500 & 49.73 \\
    500 & 500 & 32.95 \\
    500 & 1000 & 39.06 \\
    1000 & 500 & 25.98 \\
    \bottomrule
\end{tabular}
}

\newcommand{\favds}{
\begin{tabular}{ccc}
    \toprule
    \bf True Pairs & \bf False Pairs & \bf FAVD ($\downarrow$) \\
    \midrule
    \rowcolor{blue!10}
    500 & 0 & \bf 12.85 \\
    0 & 500 & 56.21 \\
    500 & 500 & 35.36 \\
    500 & 1000 & 45.17 \\
    1000 & 500 & 27.59 \\
    \bottomrule
\end{tabular}
}

\begin{table*}[!htb]
    \centering
    \begin{subtable}{0.35\textwidth}
        \centering
        \resizebox{0.85\linewidth}{!}{\favdc}
        \caption{Random selection within the same classes.}
       \label{tab: exp_favd_class}
    \end{subtable}
    \quad
    \begin{subtable}{0.35\textwidth}
        \centering
        \resizebox{0.85\linewidth}{!}{\favds}
        \caption{Random shift audio within the same pairs.}
        \label{tab: exp_favd_shift}
    \end{subtable}
    \vspace{-0.5em}
    \caption{{\bf Temporal consistency validation} of Frechet Audio-Video Distance in our \benchmark.}
        \label{tab: exp_validation_temporal}
    \vspace{-1.0em}
\end{table*}

\noindent{\textbf{Text-to-Audio Generation.}}
Text-to-audio (TTA) generation aims to develop the generative model in audio space to synthesize audio signals based on text prompts.
This task is attracting increased attention, with several works appearing in recent two years.
For instance, DiffSound~\citep{yang2022diffsound} employs a discrete diffusion process to convert audio codes from VQ-VAE into sounds. AudioGen~\citep{kreuk2023audiogen}, on the other hand, uses a two-stage process involving a neural audio compression model and an autoregressive Transformer. Additionally, Latent Diffusion Models, seen in text-to-audio systems, promise superior sound synthesis~\citep{liu2023audioldm,huang2023make}. These advanced TTA methods can now generate sounds that align with linguistic prompts with the help of cutting-edge generative models. However, a salient challenge emerges when audio is merged with video: temporal synchronization. The sounds, though of impressive quality, often don't temporally match the visual scenes. To bridge this gap, in this work, we develop a novel TTA generation method with visual alignment  based on LDMs by integrating temporal visual features with audio-visual ControlNet modules.

\section{\benchmark \& Metrics}

In this section, we propose a new benchmark, namely \benchmark, for video-aligned text-to-audio generation that can achieve both visual alignment and temporal consistency.
Furthermore, we present three novel metrics for the evaluation of the quality of our generated audio.
Finally, we also perform simulation experiments to validate the effectiveness of the proposed metrics for assessing visual alignment and temporal consistency.

\subsection{Benchmark Details}

In order to evaluate video-aligned text-to-audio generation, we develop a new dataset consisting of 500 video-text pairs.
Specifically, we downloaded all videos from 5,158 samples in the test set of VGGSound~\cite{chen2020vggsound}.
For each video, we use the class label with object and action tagging as the text caption, such as ``cat hissing'' and ``female singing''.
Based on the current video-text pairs, we curate 500 pairs by keeping videos with single source objects, temporal matching, and non-noisy backgrounds. 
With 500 video-text pairs, we extract audio clips with a length of ten seconds for each video according to the starting timestamp in the original annotated information from VGGSound~\cite{chen2020vggsound}.

\begin{figure*}[t]
    \centering
    \includegraphics[width=0.9\linewidth]{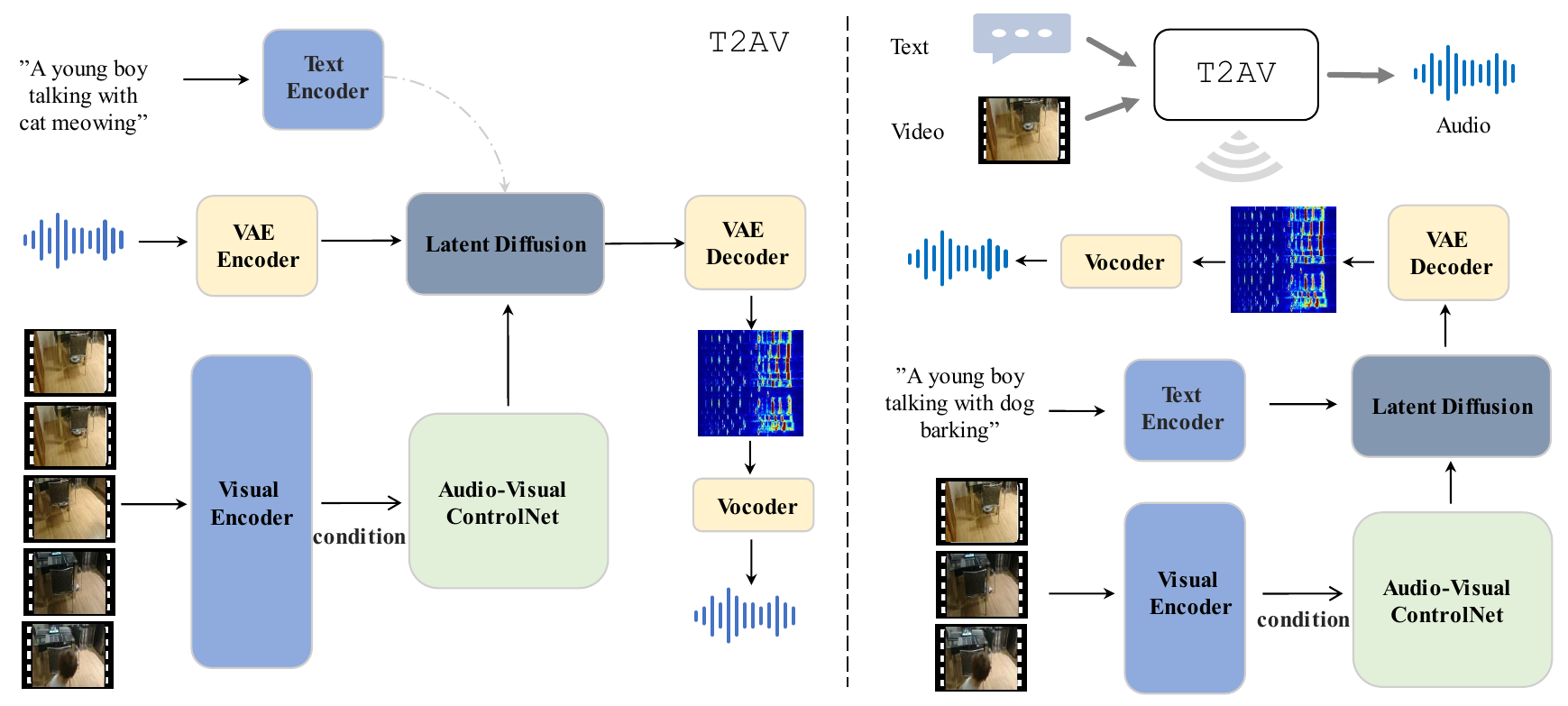}
    \vspace{-3mm}
    \caption{{\bf Illustration of the proposed framework for Text-to-audio generation aligned with videos (T2AV).}
    The Audio-Visual ControlNet aggregates temporal video features as the condition in the text-based latent diffusion models.
    Then, a contrastive language-audio pretraining objective across each temporal location is applied to match visual-aligned text embeddings with audio features. 
    After visual-aligned CLAP pre-training, we directly extract text embeddings with Audio-Visual ControlNet as the condition for latent diffusion models to achieve Text-to-Audio generation with visual alignment and temporal consistency.
    }
    \label{fig: main_img}
    \vspace{-5mm}
\end{figure*}

\subsection{Metrics Details}

Similar to Frechet inception distance (FID) to assess the quality of images and Frechet audio distance (FAD) for audio, we present three new metrics to evaluate the visual alignment and temporal consistency of generated audio in our video-aligned TTA generation.
1) {\bf Frechet Audio-Visual Distance (FAVD)}: compares the distribution of generated audio with the distribution of a set of real videos. 
Specifically, we calculate the distance between audio embeddings from VGGish~\cite{hershey2017cnn} and video embeddings from C3D~\cite{tran2015learning}.
2) {\bf Frechet Audio-Text Distance (FATD)}: evaluates the distribution of generated audio with the distribution of a set of tagging texts, where we compute the distance between audio embeddings from VGGish~\cite{hershey2017cnn} and text embeddings from word2vec~\cite{mikolov2013efficient} using the tagging class label.
3) {\bf Frechet Audio-(Video-Text) Distance (FAVTD)}: compares the distribution of generated audio with the distribution of a set of both videos and tagging texts, where we compute the distance between audio embeddings from VGGish~\cite{hershey2017cnn} and an averaged embeddings of video embeddings from VGGish~\cite{hershey2017cnn} and text embeddings from word2vec~\cite{mikolov2013efficient}.

\subsection{Metrics Validation}

In order to demonstrate the efficiency of the proposed metrics for the evaluation of visual alignment and temporal consistency, we perform simulation experiments by computing the metrics from matching (true) video-text pairs and mismatching (false) pairs.

\noindent\textbf{Visual Alignment.}
For visual alignment, we compute all metrics, including FAVD, FATD, FA(VT)D by using 500 matching (true) video-text pairs and 500 randomly selected mismatching (false) pairs in our \benchmark, and 500 matching (true) video-text pairs randomly selected from AudioCaps~\cite{kim2019audiocaps}.
The quantitative results are reported in Table~\ref{tab: exp_validation}.
With the increase in the number of false pairs with mismatching visual information, all metrics increase.
Adding 500 true pairs to [500, 500] cases further decreases all metrics.
These results validate the effectiveness of the proposed metrics in evaluating visual alignment.

\noindent\textbf{Temporal Consistency.}
For temporal consistency, we conduct two sub-experiments to compare changes in FAVD scores. First, we randomly select 500 videos from the same classes in VGGSound~\cite{chen2020vggsound}.
Second, we perform random shift audio from the same pair in our \benchmark.
Table~\ref{tab: exp_validation_temporal} shows the comparison results on our \benchmark in terms of FAVD scores.
As can be seen, the FAVD score rises with the increase in the number of false pairs with mismatching temporal consistency, although they share the same visual information.
Adding 500 additional true pairs from VGGSound~\cite{chen2020vggsound} to [500, 500] cases also decreases FAVD scores, which further shows the importance of the proposed metrics in assessing temporal consistency together with visual alignment.

\section{Method}

Given video frames and a text prompt, our aim is to synthesize an audio aligned with textual and visual semantics.
We propose a novel TTA generation approach based on LDMs personalized with visual alignment, named DiffAVA, which consists of two main modules, Visual-aligned CLAP in Section~\ref{sec:vclap} and Audio-Visual ControlNet in Section~\ref{sec:avc}.

\subsection{Preliminaries}

In this section, we first describe the problem setup and notations and then revisit conditional latent diffusion models in AudioLDM~\citep{liu2023audioldm} for TTA generation.

\noindent\textbf{Problem Setup and Notations.}
Given audio $a$ and visual frames $v$ from a video and a text prompt $t$, the goal is to generate new audio aligned with textual and visual semantics.
For a video, we have the mel-spectrogram of audio denoted as $\mathbf{A}\in\mathbb{R}^{T\times F}$, and visual frames denoted as $\mathbf{V}\in\mathbb{R}^{T\times H \times W \times 3}$. 
$T$ and $F$ denote the time and frequency, respectively.
We extract text features $\mathbf{F}^t$ and audio features $\mathbf{F}^a$ from pre-trained text encoder  $f_t(\cdot)$ and pre-trained audio encoder $f_a(\cdot)$ from CLAP~\citep{laionclap2023}.

\begin{table*}[t]
	\renewcommand\tabcolsep{6.0pt}
	\centering
	\scalebox{0.9}{
		\begin{tabular}{l|ccccccc}
			\toprule
			\bf Method & \bf FD ($\downarrow$) & \bf IS ($\uparrow$) & \bf KL ($\downarrow$) & \bf FAD ($\downarrow$) & \bf FAVD ($\downarrow$) & \bf FATD ($\downarrow$) & \bf FA(VT)D ($\downarrow$) \\ 	
			\midrule
			  SpecVQGAN~\citep{iashin2021specvqgan} & 47.72 & 3.86 & 7.53 & 9.72 & 38.12 & 23.58 & 29.16 \\
                MMDiffusion~\citep{ruan2023mmdiffusion} & 46.38 & 4.53 & 6.92 & 8.36 & 36.15 & 21.73 & 28.75 \\
                AudioGen~\citep{kreuk2023audiogen} & 45.72 & 4.61 & 6.53 & 7.82 & 35.56 & 21.36 & 28.02 \\
                AudioLDM~\citep{liu2023audioldm} & 42.59 & 4.82 & 5.72 & 6.25 & 33.76 & 19.52 & 26.37 \\
                \rowcolor{blue!10}
                \bf T2AV (ours) & \bf 33.29 & \bf 8.02 & \bf 2.12& \bf 4.05 & \bf 24.03 & \bf 13.16 & \bf 17.82 \\
			\bottomrule
			\end{tabular}}
   \vspace{-0.5em}
   \caption{{\bf Quantitative results of video-aligned text-to-audio generation on \benchmark.} Our method significantly outperforms
previous baselines in terms of all metrics.}
   \label{tab: exp_sota_t2av}
   \vspace{-5mm}
\end{table*}

\noindent\textbf{Revisit AudioLDM.}
To address the TTA generation problem, AudioLDM introduced a conditional latent diffusion model to estimate the noise $\bm{\epsilon}(\bm{z}_n, n, \mathbf{F}^t)$ from the audio prior $\bm{z}_0\in \mathbb{R}^{C\times \frac{T}{r}\times \frac{F}{r}}$ for the mel-spectrogram of audio, where $C$ and $r$ denote the channel of latent representation and the compression level, separately.
For noise estimation, they used the reweighted training objective as
\begin{equation}
    \mathcal{L}_n(\theta) = \mathbb{E}_{\bm{z}_0,\bm{\epsilon},n}\|\bm{\epsilon} - \bm{\epsilon}_\theta(\bm{z}_n, n, \mathbf{F}^a)\|
\end{equation}
where $\bm{\epsilon}\in\mathcal{N}(\bm{0},\bm{I})$ denote the added nosie.
At the final time step $N$ of the forward pass, the input $\bm{z}_n\in\mathcal{N}(\bm{0},\bm{I})$ becomes an isotropic Gaussian noise.
During the training stage, they generated the audio prior $\bm{z}_0$ from the cross-modal representation $\mathbf{F}^a$ of an audio $a$ in a video.
For TTA generation, the text embedding $\mathbf{F}^t$ is used to predict the noise $\bm{\epsilon}_\theta(\bm{z}_n, n, \mathbf{F}^t)$, instead of $\bm{\epsilon}_\theta(\bm{z}_n, n, \mathbf{F}^a)$.

\subsection{Visual-aligned CLAP}\label{sec:vclap}

To align the textual and visual features at spatial and temporal levels corresponding to the paired sound, we apply a multi-head attention transformer to aggregate temporal information from video features $\{\mathbf{F}^v_i\}_{i=1}^T$.
Then, we utilize a dual multi-modal residual network to fuse temporal visual representations with text embeddings $\mathbf{F}^t$ for generating new visual-aligned text embedding $\hat{\mathbf{F}}^t$.
Based on contrastive language-audio pre-training (CLAP), we 
apply visual-aligned CLAP between the textual features with the audio representation in the same mini-batch, which is defined as:
\begin{equation}\label{eq:m2icl}
    \mathcal{L} = 
    - \frac{1}{B}\sum_{b=1}^B \sum_{i=1}^T \log \frac{
    \exp \left( \frac{1}{\tau} \mathtt{sim}(\mathbf{F}^a_{b,i}, \hat{\mathbf{F}}^t_{b,i}) \right)
    }{
    \sum_{m=1}^B \exp \left(  \frac{1}{\tau} \mathtt{sim}(\mathbf{F}^a_{b,i}, \hat{\mathbf{F}}_{m,i}^t)\right)}
\end{equation}
where the similarity $\mathtt{sim}(\mathbf{F}^a_{b,i}, \mathbf{F}^v_{b,i})$ denotes the temporal audio-textual cosine similarity of $\mathbf{F}^a_{b,i}$ and $\mathbf{F}^v_{b,i}$ across all temporal locations at $i$-th second. 
$B$ is the batch size, $D$ is the dimension size, and $\tau$ is a temperature hyper-parameter.

\subsection{Audio-Visual ControlNet}\label{sec:avc}

With the benefit of visual-aligned CLAP pre-training, we use the pre-trained text encoder to extract text embeddings $\hat{\mathbf{F}}^t$ with visual-aligned semantics as the condition for latent diffusion models.
In order to enhance the temporal consistency, we are inspired by ControlNet~\cite{zhang2023adding} for multimodal conditional text-to-image diffusion models and propose a novel Audio-Visual ControlNet module with temporal self-attention layers, as shown in Figure~\ref{fig: main_img}.
Specifically, we apply temporal self-attention layers $\phi(\cdot)$ to aggregate temporal features from the raw output features of the pre-trained visual encoder as:
\begin{equation}
\begin{aligned}
    \{\hat{\mathbf{F}}_{i}^v\}_{i=1}^T = \{\phi(\mathbf{F}_i^v, \mathbf{F}^v, \mathbf{F}^v)\}_{i=1}^{T},\quad 
    \mathbf{F}^v = \{\mathbf{F}_i^v\}_{i=1}^T
\end{aligned}
\end{equation}
The self-attention operator $\phi(\cdot)$ is formulated as:
\begin{equation}
    \phi(\mathbf{F}_i, \mathbf{F}^v, \mathbf{F}^v) = \mbox{Softmax}(\dfrac{\mathbf{F}_i(\mathbf{F}^v)^\top}{\sqrt{D}})\mathbf{\mathbf{F}^v}
\end{equation}
where $[\ ;\ ]$ denotes the concatenation operator.
$\mathbf{F}_i^v\in\mathbb{R}^{1\times D}$, and $D$ is the dimension of embeddings. 
Note that our model supports to freezing of the parameters of latent diffusion models and directly uses the VAE and vocoder released from AudioLDM to achieve efficient and video-conditioned TTA generation.

\begin{table}[t]
	\renewcommand\tabcolsep{6.0pt}
	\centering
	\scalebox{0.78}{
		\begin{tabular}{l|cccc}
			\toprule
			\bf Method & \bf IS ($\uparrow$) & \bf KL ($\downarrow$) & \bf FAD ($\downarrow$) & \bf FD ($\downarrow$)  \\ 	
			\midrule
			DiffSound~\citep{yang2022diffsound} & 4.01 & 2.52 & 7.75 & 47.68 \\
                AudioGen~\citep{kreuk2023audiogen} & -- & 2.09 & 3.13 & -- \\
                AudioLDM~\citep{liu2023audioldm} & 6.90 & 1.97 & 2.43 & 29.48 \\	
                \rowcolor{blue!10}
                \bf T2AV (ours) & \bf 8.25 & \bf 1.32 & \bf 1.78 & \bf 25.38 \\
			\bottomrule
			\end{tabular}}
   \vspace{-0.5em}
   \caption{{\bf Quantitative results of text-to-audio generation on AudioCaps benchmark.}  We achieve the best results in terms of all metrics on text-to-audio generation.}
   \label{tab: exp_sota_audiocaps}
   \vspace{-7mm}
\end{table}

\begin{figure*}[t]
\centering
\includegraphics[width=0.99\linewidth]{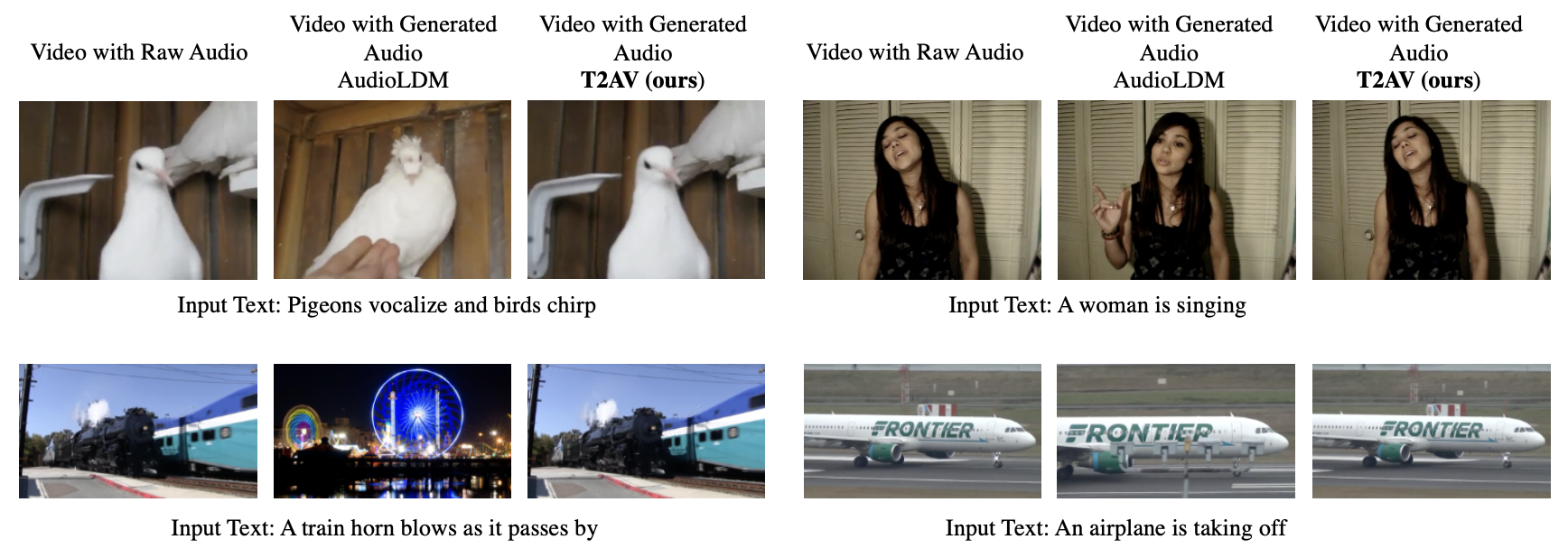}
\vspace{-1.0em}
\caption{{\bf Qualitative comparisons with AudioLDM on video-aligned TTA generation.} 
The proposed T2AV produces more accurate and aligned audio for target videos. 
}
\label{fig: vis_cmp}
\vspace{-3mm}
\end{figure*}

\begin{table*}[t]
	\renewcommand\tabcolsep{6.0pt}
	\centering
	\scalebox{0.9}{
		\begin{tabular}{ccccccccc}
			\toprule
			\bf VCLAP & \bf AVCN & \bf FD ($\downarrow$) & \bf IS ($\uparrow$) & \bf KL ($\downarrow$) & \bf FAD ($\downarrow$) & \bf FAVD ($\downarrow$) & \bf FATD ($\downarrow$) & \bf FA(VT)D ($\downarrow$) \\ 	
			\midrule
               \xmark & \xmark & 42.59 & 4.82 & 5.72 & 6.25 & 33.76 & 19.52 & 26.37 \\
                \cmark & \xmark & 37.28 & 7.53 & 3.45 & 5.52 & 29.63 & 16.93 & 23.02 \\
                \xmark & \cmark & 38.32 & 6.85 & 3.57 & 5.65 & 30.35 & 18.05 & 24.53 \\
                \rowcolor{blue!10}
			  \cmark & \cmark & \bf 33.73 & \bf 8.31 & \bf 2.62	& \bf 4.26 & \bf 24.15 & \bf 13.65 & \bf 18.05 \\
			\bottomrule
			\end{tabular}}
    \vspace{-0.5em}
   \caption{{\bf Ablation studies on Visual-aligned CLAP (VCLAP) and Audio-Visual ControlNet (AVCN).} 
   All models are trained on AudioCaps and VGGSound, and the parameters of latent diffusion models are also {\bf trainable}.}
   \label{tab: exp_ablation}
   \vspace{-3mm}
\end{table*}

\begin{table*}[t]
	\renewcommand\tabcolsep{6.0pt}
	\centering
	\scalebox{0.85}{
		\begin{tabular}{cccccccc}
			\toprule
			\bf Condition Operators & \bf FD ($\downarrow$) & \bf IS ($\uparrow$) & \bf KL ($\downarrow$) & \bf FAD ($\downarrow$) &  \bf FAVD ($\downarrow$) & \bf FATD ($\downarrow$) & \bf FA(VT)D ($\downarrow$) \\ 	
			\midrule
                LSTM & 40.78 & 4.63 & 4.86 & 6.02 & 30.57 & 18.93 & 24.85 \\
                Cross-Attention & 38.73 & 6.51 & 3.83 & 5.82 & 30.73 & 18.36 & 24.92 \\
                Addition & 37.56 & 7.32 & 3.67 & 5.79 & 29.85 & 17.19 & 23.29 \\ 
                \rowcolor{blue!10}
                \bf Temporal Self-Attention & \bf 36.98 & \bf 7.52 & \bf 3.52 & \bf 5.61 & \bf 28.65 & \bf 16.95 & \bf 22.68 \\
			\bottomrule
			\end{tabular}}
   \vspace{-0.5em}
   \caption{{\bf Exploration studies on condition operators in Audio-Visual ControlNet.} All models are trained on AudioCaps and VGGSound, and the parameters of latent diffusion models are {\bf frozen}.}
   \label{tab: ab_condition}
   \vspace{-5mm}
\end{table*}

\section{Experiments}

\subsection{Experimental setup}

\noindent \textbf{Datasets.}
AudioCaps~\cite{kim2019audiocaps} dataset includes 45,423 ten-second audio clips collected from YouTube videos paired with captions for training and 2,240 samples for validation.
Since each audio clip in AudioCaps has 5 text captions, we use the same testing set in AudioLDM~\cite{liu2023audioldm} with 886 instances by selecting one random caption as a text condition.
VGGSound~\cite{chen2020vggsound} contains 200k YouTube video clips of
10 seconds long from 309 sound categories, such as such as animals, vehicles, human speech, dancing, musical instruments, etc.
The Look, Listen and Parse (LLP) Dataset~\cite{tian2020unified} includes 11,849 YouTube video clips of
10 seconds long from 25 different event classes, such as music, car, etc.

\noindent \textbf{Evaluation Metrics.}
For comprehensive evaluation between generated audio and target audio, we apply Inception Score (IS), Kullback–Leibler (KL) divergence, Frechet Audio Distance (FAD), and Frechet Distance (FD) as evaluation metrics, following the previous work~\cite{liu2023audioldm}.
IS is used to measure both audio quality and diversity, while KL is evaluated on paired audio.
FAD and FD calculate the similarity between generated audio and reference audio. 
For video-aligned TTA generation, we use the proposed metrics, including FAVD, FATD, FA(VT)D for evaluation.

\noindent \textbf{Implementation.}
We initialize the weights from the audio and text encoder in AudioLDM~\cite{liu2023audioldm} and fine-tune the parameters. 
For the video encoder, we apply the pre-trained X-CLIP~\cite{ma2022xclip} as the initialization weights.
The depth of multi-head self-attention layers in Audio-Visual ControlNet with a dimension of 768 is 4, and the number of heads is 8.
The model is trained for 50 epochs using a batch size of 128 and the Adam optimizer with a learning rate of $1e-4$.
We use the released weights from AudioLDM~\cite{liu2023audioldm} for VAE and vocoder to generate the final audio samples.

\subsection{Comparison to Prior Work}\label{sec:exp}

In this work, we propose a novel and effective framework for text-to-audio generation. 
In order to validate the effectiveness of the proposed T2AV, we comprehensively compare it to previous DDPM and LDM baselines:
1) SpecVQGAN~\cite{iashin2021specvqgan} (2021'BMVC): a VQGAN approach based on a compact sampling space to generate a new spectrogram from the pre-trained spectrogram codebook.
2) DiffSound~\cite{yang2022diffsound} (2022'TASLP): a vector-quantized variational autoencoder (VQ-VAE) based DDPM framework by learning a discrete space from audio given natural language description with mask-based text generation.
3) MMDiffusion~\cite{ruan2023mmdiffusion} (2023'CVPR): a strong multi-modal diffusion model with a sequential U-Net with two-coupled autoencoders for a joint audio-video denoising process.
4) AudioGen~\cite{kreuk2023audiogen} (2023'ICLR):
a recent DDPM approach using a transformer decoder to learn discrete representations from the audio waveform directly. 
5) AudioLDM~\cite{liu2023audioldm} (2023'ICML):
a strong LDM baseline to learn the continuous audio representations from a latent space in contrastive language-audio pre-training.

For video-aligned text-to-audio generation, we report the quantitative comparison results on the \benchmark in Table~\ref{tab: exp_sota_t2av}.
As can be seen, we achieve the best results regarding all metrics for video-aligned text-to-audio generation compared to previous DDPM and LDM approaches.
In particular, the proposed T2AV superiorly outperforms AudioLDM~\citep{liu2023audioldm}, the current state-of-the-art text-to-audio generation baseline, by 9.30 FD \& 2.20 FAD, 9.73 FAVD \& 6.36 FATD, and 8.55 FA(VT)D on our benchmark.
Furthermore, we achieve significant performance gains compared to MMDiffusion~\citep{ruan2023mmdiffusion}, the strong multi-modal joint diffusion baseline, which indicates the importance of extracting temporal visual semantics from visual representations as guidance for video-aligned text-to-audio generation.
Meanwhile, our  achieves better results against
those DDPM baselines, such as DiffSound~\cite{yang2022diffsound} and AudioGen~\cite{kreuk2023audiogen}.
These significant improvements demonstrate the superiority of our approach in generating high-quality video-aligned audio from text captions.

In addition, significant gains in TTA generation on the AudioCaps benchmark can be observed in Table~\ref{tab: exp_sota_audiocaps}.
As can be seen, we achieve the best results in terms of IS and KL while performing competitively in other metrics.
In particular, the proposed  significantly outperforms DiffSound~\cite{yang2022diffsound}, the first DDPM-based baseline on TTA generation, by 4.24 IS, and highly decreases other metrics by 1.20 KL, 5.97 FAD, and 22.30 FD.
Moreover, we achieve decent performance gains of 1.35 IS and a decrease of 4.10 FD,  compared to AudioLDM~\cite{liu2023audioldm}, the current state-of-the-art text-to-audio generation approach.
These results demonstrate the effectiveness of our approach in learning visual-aligned textual semantics for text-to-audio generation.

In order to qualitatively evaluate the quality of video-aligned TTA generation, we compare the proposed  with AudioLDM~\cite{liu2023audioldm} in Figure~\ref{fig: vis_cmp}.
From comparisons, we observed that without explicit visual-aligned CLAP pre-training objectives, AudioLDM~\cite{liu2023audioldm}, the strong TTA generation baseline fails to discriminate the order of two sounding objects given in the input text, such as ``Pigeons vocalize and birds chirp''.
Meanwhile, it is hard for AudioLDM~\cite{liu2023audioldm} to generate sound temporally aligned with the original video.
For example, given the input text ``A train horn blows as it passes by'', the strong baseline model generates the horn blowing sound across all ten seconds although no train appeared in the last several seconds. 
In contrast, the generated audio from our method is more aligned with the visual semantics existing in the video.
These visualizations further showcase the superiority of our approach in video-aligned text-to-audio generation.

\subsection{Experimental Analysis}

In this section, we provide ablation studies to demonstrate the benefit of the Visual-aligned CLAP and Audio-Visual ControlNet modules.
We also conducted extensive experiments to explore the condition in Audio-Visual ControlNet, training data scale, and latent diffusion tuning.

\begin{table*}[t]
	\renewcommand\tabcolsep{6.0pt}
	\centering
	\scalebox{0.85}{
		\begin{tabular}{cccccccc}
			\toprule
			\bf Train Data & \bf FD ($\downarrow$) & \bf IS ($\uparrow$) & \bf KL ($\downarrow$) & \bf FAD ($\downarrow$) & \bf FAVD ($\downarrow$) & \bf FATD ($\downarrow$) & \bf FA(VT)D ($\downarrow$) \\ 	
			\midrule
                AudioCaps & 36.98 & 7.52 & 3.52 & 5.61 & 28.65 & 16.95 & 22.68 \\
                AudioCaps + VGGSound & 35.28 & 8.72 & 2.93 & 4.78 & 25.23 & 14.71 & 20.16 \\
                \rowcolor{blue!10}
                \bf AudioCaps + VGGSound + LLP & \bf 34.86 & \bf 9.26 & \bf 2.35 & \bf 4.21 & \bf 24.15 & \bf 13.32 & \bf 18.95 \\
			\bottomrule
			\end{tabular}}
   \vspace{-0.5em}
   \caption{{\bf Exploration studies on the scale of training data.} 
   The parameters of latent diffusion models are frozen.}
   \label{tab: ab_data}
   \vspace{-1mm}
\end{table*}

\begin{table*}[t]
	\renewcommand\tabcolsep{6.0pt}
	\centering
	\scalebox{0.85}{
		\begin{tabular}{ccccccccc}
			\toprule
			\bf Tuning & \bf Train Data & \bf FD ($\downarrow$) & \bf IS ($\uparrow$) & \bf KL ($\downarrow$) & \bf FAD ($\downarrow$) & \bf FAVD ($\downarrow$) & \bf FATD ($\downarrow$) & \bf FA(VT)D ($\downarrow$) \\ 	
			\midrule
                \xmark & AC + VS & 35.28 & 8.72 & 2.93 & 4.78 & 25.23 & 14.71 & 20.16 \\
                \cmark & AC + VS & \bf 33.73 & \bf 8.31 & \bf 2.62 & \bf 4.26 & \bf 24.15 & \bf 13.65 & \bf 18.05 \\ \hline
                \xmark & AC + VS + LLP & 34.86 & 9.26 & 2.35 & 4.21 & 24.15 & 13.32 & 18.95 \\
                \rowcolor{blue!10}
                \cmark & AC + VS + LLP & \bf 33.29 & \bf 8.02 & \bf 2.12 & \bf 4.05 & \bf 24.03 & \bf 13.16 & \bf 17.82 \\
			\bottomrule
			\end{tabular}}
   \vspace{-0.5em}
   \caption{{\bf Exploration studies on the latent diffusion tuning.} AC and VS denote AudioCaps and VGGSound datasets, respectively.
   All models are trained on
AC + VS or AC + VS + LLP, and the parameters of latent diffusion models are also trainable.}
   \label{tab: ab_tune}
   \vspace{-5mm}
\end{table*}

\noindent\textbf{Visual-aligned CLAP \& Audio-Visual ControlNet.}
In order to demonstrate the effectiveness of the introduced visual-aligned contrastive language-audio pre-training (VCLAP) and audio-visual ControlNet (AVCN), we ablate the necessity of each module and report the quantitative results on our \benchmark in Table~\ref{tab: exp_ablation}.
We train our model on AudioCaps and VGGSound, and also tune the parameters of latent diffusion models for a comprehensive comparison.
As can be observed, adding VCLAP to the vanilla baseline highly decreases the performance by 5.31 FD \& 0.73 FAD, 4.13 FAVD \& 2.59 FATD, and 3.35 FA(VT)D, which validates the benefit of visual-aligned contrastive language-audio pre-training in extracting text representations with visual semantics for text-to-audio generation. 
Meanwhile, introducing only AVCN in the baseline increases the video-aligned text-to-video generation performance regarding all metrics.
More importantly, incorporating VCLAP and AVCN into the baseline significantly reduces the results of 9.30 FD \& 2.20 FAD, 9.73 FAVD \& 6.36 FATD, and 8.55 FA(VT)D on our benchmark. 
These improving results validate the importance of visual-aligned contrastive language-audio pre-training and audio-visual ControlNet in extracting temporal-aware semantics from videos as guidance for text-to-audio generation.

\noindent\textbf{Condition in Audio-Visual ControlNet.}
Learning video-aligned textual representations with temporal-aware semantics as the condition in the proposed Audio-Visual ControlNet module is critical for generating high-quality audio from input captions.
To explore such effects more comprehensively, we varied the condition operator from $\{$LSTM, Cross-Attention, Addition, Temporal Self-Attention$\}$.
The comparison results of video-aligned text-to-audio generation performance are reported in Table~\ref{tab: ab_condition}.
When using the vanilla LSTM as the condition operator, we achieve the worst results in terms of all metrics. 
With a temporal self-attention operator to aggregate temporal-aware visual features as guidance, the proposed T2AV achieves the best video-aligned text-to-audio generation performance in terms of all metrics decreasing by 3.80 FD \& 0.41 FAD, 1.92 FAVD \& 1.98 FATD, and 2.17 FA(VT)D, which shows the importance of long-term temporal features in Audio-Visual ControlNet for generating high-fidelity audio aligned with videos.
Regarding the Cross-Attention operator, the performance of the proposed T2AV performs better than the vanilla LSTM decreasing by 2.05 FD \& 0.20 FAD, 0.16 FAVD \& 0.57 FATD, and 0.07 FA(VT)D when only global visual features are fed into latent diffusion models as guidance. 
Interestingly, using a simple addition operator on top of temporal video features will continually improve the result against the vanilla LSTM by decreasing 3.22 FD \& 0.23 FAD, 0.72 FAVD \& 1.74 FATD, and 1.56 FA(VT)D since there might be some temporal consistency maintained in the original visual features from the video encoder by visual-aligned contrastive language-audio pre-training to boost the performance of video-aligned text-to-audio generation.

\noindent\textbf{Training Data Scale.}
In order to show the scaling-up properties of our T2AV in large-scale training data, we ablated the training data from $\{$AudioCaps, AudioCaps+VGGSound, AudioCaps+VGGSound+LLP$\}$ and report the comparison results on our \benchmark in Table~\ref{tab: ab_data}.
Note that all models are trained except for the parameters of latent diffusion models are frozen.
As can be seen, we achieve the worst results on our \benchmark in terms of all metrics, when using only AudioCaps~\cite{kim2019audiocaps} for training.
By adding VGGSound~\cite{chen2020vggsound} with 200k videos to AudioCaps in the training data, we achieve better video-aligned text-to-audio generation performance in terms of all metrics decreasing by 1.70 FD \& 0.83 FAD, 3.42 FAVD \& 2.24 FATD, and 2.52 FA(VT)D, which shows the importance of diverse categories in VGGSound in learning video-aligned representations for generating high-fidelity audio.
Furthermore, incorporating LLP~\cite{tian2020unified} into the previous two training datasets continually improves the performance by decreasing 2.12 FD \& 1.40 FAD, 4.50 FAVD \& 3.63 FATD, and 3.73 FA(VT)D, compared to the vanilla baseline trained on only AudioCaps.
This might be due to the rich temporal semantics in LLP training data as each video includes at least one-second audio or visual events, which also validates the importance of learning temporal semantics from the video encoder by visual-aligned contrastive language-audio pre-training in video-aligned text-to-audio generation.

\noindent\textbf{Latent Diffusion Tuning.}
Tuning the latent diffusion model to capture temporal-aware representations as the guidance also benefits generating high-quality videos from input captions.
In order to explore such effects more comprehensively on video-aligned text-to-audio generation, we ablated latent diffusion tuning and varied the training data from $\{$AC + VS, AC + VS + LLP$\}$, where AC and VS denote AudioCaps and VGGSound datasets, respectively.
The comparison results on our \benchmark are shown in Table~\ref{tab: ab_tune}.
Adding latent diffusion model tuning to both training data settings improves video-aligned text-to-audio generation performance in terms of all metrics.
In particular, latent diffusion tuning improves the results by decreasing 1.55 FD \& 0.52 FAD, 1.08 FAVD \& 1.06 FATD, and 2.11 FA(VT)D on AC + VS, and by 1.57 FD \& 0.16 FAD, 0.12 FAVD \& 0.16 FATD, and 1.13 FA(VT)D on AC + VS + LLP.
These improvements demonstrate the importance of latent diffusion tuning in learning new temporal-aware semantics aligned with both text and videos using Audio-Visual ControlNet for generating high-fidelity audio aligned with videos from the input caption.
However, the performance gains of latent diffusion model tuning on AC + VS + LLP are less than latent diffusion model tuning on AC + VS, since there might be some similar videos with overlapping semantics from the video encoder by visual-aligned contrastive language-audio pre-training on AC + LLP or VS + LLP, which leads to confusion on generating audio aligned with different videos from similar input captions.

\section{Conclusion}

In this work, we present \benchmark, a new benchmark for TTA generation aligned with videos, and three novel metrics that evaluate visual alignment and temporal consistency.
We also propose a simple yet effective latent diffusion approach, named \method, that integrates temporal visual-aligned embeddings with Audio-Visual ControlNet as the condition. 
We introduce a temporal multi-head self-attention transformer to aggregate temporal information from video features to fuse temporal visual embeddings in Audio-Visual ControlNet.
Empirical experiments on AudioCaps and our \benchmark benchmarks demonstrate the state-of-the-art performance of our \method on visual-aligned text-to-audio generation.
Meanwhile, qualitative visualizations of videos paired with audio vividly showcase the effectiveness of our \method in capturing visual alignment and temporal consistency.
Furthermore, extensive ablation studies also validate the importance of visual-aligned CLAP and Audio-Visual ControlNet, training data scale, and latent diffusion tuning in video-aligned text-to-audio generation.

\section*{Impact Statement}
The method under discussion creates audio that aligns with videos based on user-uploaded captions on the web. 
This approach risks embedding internal biases in the data. 
For instance, the model may struggle to accurately produce audio for less common yet vital sound categories in videos. 
Addressing these potential shortcomings is crucial for the application in real-world scenarios.

\bibliography{reference}
\bibliographystyle{icml2024}


\end{document}